\documentclass[aps,prl,twocolumn,showpacs]{revtex4}
\usepackage{epsfig}
\usepackage{times}
\begin{document}

\title{Cooperation enhanced by inhomogeneous activity of teaching for evolutionary Prisoner's Dilemma games}

\author{Attila Szolnoki and Gy\"orgy Szab\'o}
\affiliation
{Research Institute for Technical Physics and Materials Science,
P.O. Box 49, H-1525 Budapest, Hungary}

\date{\today}

\begin{abstract}
Evolutionary Prisoner's Dilemma games with quenched inhomogeneities in the spatial dynamical rules are 
considered. The players following one of the two pure strategies (cooperation or defection) are distributed 
on a two-dimensional lattice. The rate of strategy adoption from a randomly chosen neighbors are controlled 
by the payoff difference and a two-value pre-factor $w$ characterizing the players whom the strategy learned 
from. The reduced teaching activity of players is distributed randomly with concentrations $\nu$ at the 
beginning and fixed further on. Numerical and analytical calculations are performed to study the 
concentration of cooperators as a function of $w$ and $\nu$ for different noise levels and connectivity 
structures. Significant increase of cooperation is found within a wide range of parameters for this dynamics. 
The results highlight the importance of asymmetry characterizing the exchange of master-follower role during 
the strategy adoptions.
\end{abstract}

\pacs{02.50.+Le, 87.23.Ge, 07.05.Tp}

\maketitle

Different versions of evolutionary Prisoner's Dilemma (PD) games are studied extensively to explore the 
possibilities enhancing the cooperative (altruistic) behavior among selfish individuals. Originally the 
Prisoner's Dilemma represents those class of two-player symmetric matrix games where the equivalent players 
have two choices [called cooperation ($C$) or defection ($D$)] with a specific rank of order in the four 
payoffs (dependent on their choices) enforcing both selfish (rational) individuals to choose defection 
instead of mutual cooperation yielding significantly higher individual incomes for them 
\cite{weibull_95,gintis_00,cressman_03}. This type of social dilemma cannot be resolved within the framework 
of the traditional (two-player, one-shot) game theory. During the last decades different mechanisms, e.g., 
kin selection \cite{hamilton_jtb64a}, direct \cite{axelrod_s81} and indirect \cite{fehr_n02} reciprocity, 
voluntary participation \cite{hauert_s02}, and spatial extensions \cite{nowak_ijbc93}, are found to support 
the emergence of cooperative behavior in biological and ecological systems as well as within human societies 
\cite{nowak_s04}.  

In a wide class of the spatial models \cite{nowak_ijbc93,lindgren_pd94,nakamaru_jtb97}, the 
players distributed on the sites $x$ of a lattice can follow one of the pure strategies ($s_x=C$ or $D$), 
their payoffs $U_x$ come from PD games with their neighbors, and sometimes the players are allowed to modify 
their strategy according to an evolutionary rule dependent on the local payoff distribution. The analogy between these spatial evolutionary games and the non-equilibrium kinetic Ising models has motivated a progressive interest to utilize the tools of non-equilibrium statistical physics in the investigation of evolutionary games (for a review see \cite{szabo_cm06}). In these models the connectivity structure of the interacting players is described by a graph represented by sites (players) and edges between the neighbors. Numerical investigations on realistic connectivity structures (e.g., diluted \cite{nowak_ijbc94,vainstein_pre01} and hierarchical lattice \cite{vukov_pre05}, random graphs \cite{ebel_pre02,duran_pd05}, small-world structures \cite{abramson_pre01,wu_cpl06}, and real empirical networks \cite{holme_pre03}) have also been performed for some time. The early investigations have indicated some increase in the concentration of cooperators when the topological inhomogeneities are increased. Similar phenomena can also be observed if the players are allowed to move on a lattice or graph (for a recent survey see \cite{vainstein_cm06}). The systematic investigations, however, are delayed by the large number of parameters characterizing the payoffs, the connectivity structures, and the evolutionary rules that can involve synchronization and/or noise and be affected by the connectivity structure too.

In the last year Santos {\it et al.} \cite{santos_prl05,santos_prslb06} observed dramatic improvement in the maintenance of cooperation when considering an evolutionary PD game on scale-free structures at a low noise level. In these systems the competition between the cooperative and defective hubs (players with large number of neighbors) provides a mechanism \cite{santos_prslb06} enforcing cooperation so much so that defectors can die out. The given evolutionary rule involves intrinsically an inhomogeneous strategy adoption probability due to the varying number of neighbors. Remarkable increase of cooperation is also observed for those systems where the inhomogeneous imitation activity is introduced artificially to characterize the asymmetric and different influence of players to each other \cite{kim_pre02}. In the models suggested by Wu {\it et al.} \cite{wu_cpl06,wu_pre06}, the influence of player $y$ to her neighbor $x$ is quantified by a random parameter $w_{xy}$ ($0 < w_{xy} \ne w_{yx} < 1 $) affecting the preferential selection of a neighbor whom the strategy can be adopted from. 

Now we study the effect of payoffs and noise on the maintenance of cooperative behavior for some evolutionary PD games where the inhomogeneities are involved in the dynamical rules on regular connectivity structures. Our motivation was to explore the consequence of varying activity in the strategy imitation process. For this purpose, a simplified version of the model introduced by Wu {\it et al.} \cite{wu_cpl06,wu_pre06} will be considered by assuming only two possible values for $w_{xy}$. First our investigation is focused on a system where $w_{xy}$ depends only on $y$ and characterizes the teaching (helping) activity of player $y$ during the strategy imitations.

In the first model two types of players ($A$ and $B$) are distributed randomly on a two-dimensional lattice before the start of simulation and their distribution ($n_x=A$ or $B$) is fixed later on. The concentration of players $B$ and $A$ are denoted by $\nu$ and $(1-\nu)$. Both types of players can follow the $C$ or $D$ strategies and their total payoff comes from a PD game with the neighbors as formulated in previous papers (see e.g., \cite{szabo_pre05}). For a given two-player game both players receive reward $R$ (or punishment $P$) for mutual cooperation (or defection). If the players follow different strategies then the defector receives the highest payoff $T$ (temptation to choose defection) meanwhile the cooperator gets the lowest (sucker's) payoff $S$. Henceforth our analysis will be restricted to the parametrization suggested by Nowak and May \cite{nowak_ijbc93}, i.e., $R=1$; $P=0$; $T=b$ ($1 < b < 2$); and $S=0$.

For the evolutionary PD games the players try to maximize their individual payoff by imitating (learning) one of the more successful neighboring strategies. Following our previous studies \cite{szabo_pre05,szabo_cm06}, the evolution of the present system is governed by subsequent strategy adoptions between randomly chosen neighbors $x$ and $y$. Namely, player $x$ will adopt the neighbor's strategy $s_y$ with a probability depending on the payoff difference $(U_x-U_y)$ as 
\begin{equation}
\label{eq:Wtotal}
W(s_x \to s_y)=w_{xy}{1 \over 1 + \exp [(U_x-U_y)/K] }
\end{equation}
where $K$ denotes the amplitude of noise. The pre-factor $w_{xy}$ is given as
\begin{equation}
\label{eq:wobt}
w_{xy}= \cases {1, &if $n_y=A$ \cr
                w, &if $n_y=B$ \cr }\; ,
\end{equation}
where the value of $w$ ($0< w < 1$) characterizes the strength of reduced teaching activity if the site $y$ is occupied by a player of type $B$. One can think of a system consisting of attractive and repulsive players (just like old and young individuals in some communities). 

\begin{figure}[ht]
\centerline{\epsfig{file=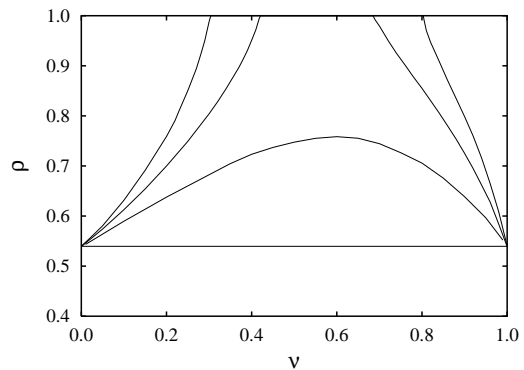,width=7cm}}
\caption{\label{fig:rb}The concentration of cooperators ($\rho$) as a function of $\nu$ for different rates 
of blocking (from bottom to top, $w=1, 0.2, 0.1$, and $0.05$) on the Kagome lattice if $b=1.03$ and $K=0.5$. 
Due to the large system size ($3 \cdot 10^5 < N < 3 \cdot 10^6$) and long sampling times the statistical 
errors of MC data are comparable to the line thickness.}
\end{figure}

This system is studied by Monte Carlo (MC) simulations started from a random initial distribution of $C$ and $D$ strategies. The stationary state is characterized by the average concentration $\rho$ of cooperators when the values of $K$, $b$, and $\nu$ are varied systematically. First we discuss the MC results on the Kagome lattice. The striking increase of $\rho$ as a result of this type of dynamical inhomogeneities is illustrated in Fig.~\ref{fig:rb}.

Evidently, this model is equivalent to a homogeneous system (discussed in \cite{szabo_pre05}) if $w=1$. 
Furthermore, the average concentration of $C$ strategies for $\nu=0$ is equivalent to those for $\nu = 1$ 
because the dynamics is homogeneous in both cases (the relaxation, however, is slower for $\nu=1$ if $w<1$). 
The value of $\rho$ increases monotonously until reaching the maximum value at $\nu \approx 0.5$. For 
sufficiently small values of $w$ the concentration of cooperators reaches its saturation value. 
Figure~\ref{fig:rb} shows clearly the existence of a parameter range within which the defectors become 
extinct in the stationary states (even for $b>1$). The rigorous numerical analysis of the transition from 
the two-strategy ($C+D$) state to the homogeneous (absorbing) state is made difficult by the slow relaxation 
processes characterizing those systems where "directed percolation" type extinction process \cite{marro_99} 
is disturbed by quenched disorder in the background \cite{dickman_pre98,hooyberghs_pre04}. 

\begin{figure}[ht]
\centerline{\epsfig{file=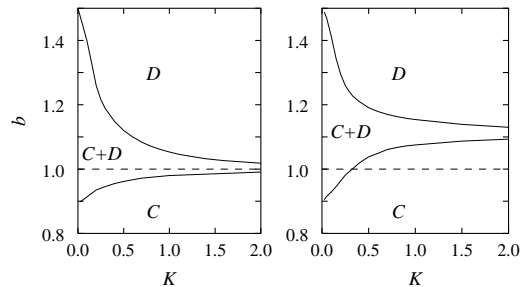,width=7cm}}
\caption{\label{fig:phds}The complete $b-K$ phase diagrams for the Prisoner's Dilemma game on the Kagom\'e lattice for $\nu=0$ (left) and $0.5$ (right). Dashed lines indicate the common limit value of $b_{c1}$ and $b_{c2}$ when $K \to \infty$ in the homogeneous system.}
\end{figure}

Now the systematic study is addressed to explore the variations in the $b-K$ phase diagram caused by the inhomogeneous dynamics. For this purpose we have determined the critical values ($b_{c1}$ and $b_{c2}$) where the extinction of $D$ and $C$ strategies occur when varying $b$ for different noise levels ($K$). Figure \ref{fig:phds} compares two phase diagrams obtained for $\nu = 0$ and $\nu = 0.5$ (at $w=0.1$).

\begin{figure}[ht]
\centerline{\epsfig{file=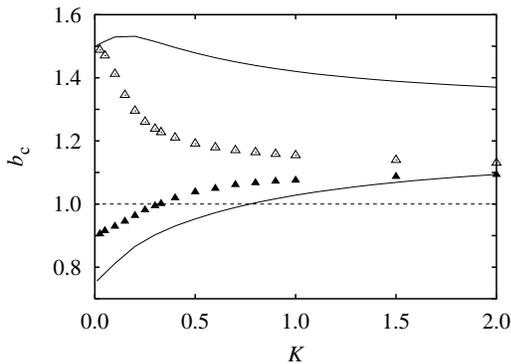,width=7cm}}
\caption{\label{fig:lsk3pmc}Phase boundaries for the evolutionary PD game on the Kagom\'e lattice for $\nu=0.5$ and $w=0.1$. Solid lines show the predictions of the three-site approximation and the symbols indicate the MC results.}
\end{figure}

Figure \ref{fig:phds} illustrates clearly the remarkable difference appearing in the high noise limit ($K \to \infty$) while the effect of quenched inhomogeneities vanishes if $K \to 0$. The mean-field calculation \cite{szabo_cm06} predicts a step-like change between the homogeneous $C$ and $D$ states at $b=1$. The simulations reproduce this prediction for only the homogeneous system in the limit $K \to \infty$. According to the simulations the phase boundaries ($b_{c1}$ and $b_{c2}$) tend towards the same limit $b_{\infty}$ (if $K \to \infty$), however, the corresponding limit value is larger than 1. The increase of $b_{\infty}$ (dependent on $w$ and $\nu$) cannot be described by the mean-field approach. To overcome this shortage we have determined the probability of all possible configurations on a triangle of sites allowing the uncorrelated distribution of players $A$ and $B$ (survey of this technique is given in the Appendix of \cite{szabo_cm06}). Evaluating $4^3$ configuration probabilities this approach is capable to reproduce qualitatively the relevant features as demonstrated in Fig.~\ref{fig:lsk3pmc}.

In the low noise limit the maintenance of cooperation on the Kagome lattice (for $b>1$) is due to a mechanism 
supporting the spreading of cooperators throughout overlapping triangles of the connectivity structure 
\cite{szabo_pre05,vukov_pre06}. In the absence of this topological feature $b_{c2}$ goes to 1 linearly 
if $K \to 0$. For example, on the square lattice the cooperation disappear in these types of evolutionary 
PG games if $K \to 0$. When increasing the noise level one can observe a peak in the concentration of 
cooperators analogously to the so-called coherence resonance \cite{traulsen_prl04,perc_njp06}. The 
significant differences in the low noise behavior allows us to check the robustness of the observed 
phenomenon. For this purpose the above analysis is repeated on the square lattice.

\begin{figure}[ht]
\centerline{\epsfig{file=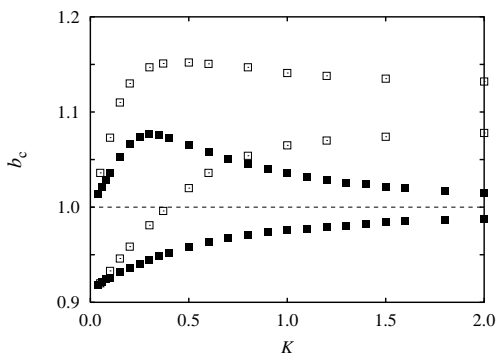,width=7cm}}
\caption{\label{fig:lssqrmc}Phase diagram of the evolutionary PD game on the square lattice. 
Closed (open) squares indicate MC results for $\nu=0$ ($\nu=0.5$).}
\end{figure}

The MC results on the square lattice (see Fig.~\ref{fig:lssqrmc}) illustrates the same trend in the variation of $b_{c1}$ and $b_{c2}$ when the quenched inhomogeneity of dynamics is applied ($w=0.1$ and $\nu = 0.5$).

The plotted results confirm the above mentioned conclusions. Namely, the effect of inhomogeneous dynamics vanishes in the zero noise limit and increases the values of $b_{c1}$ and $b_{c2}$. It is worth mentioning that the pair approximation fails for the homogeneous system in the the low noise limit, while its prediction is qualitatively correct for large $K$ values \cite{szabo_pre05}. In agreement with the expectation, the extended version of the (homogeneous) pair approximation gives account of the simultaneous shift of both phase boundaries in a way that $(b_{c2}-b_{c1}) \to 0$ if $K \to \infty$. The extended method takes explicitly into account the different concentrations of cooperators on the sites of type $A$ and $B$ as well as distinct correlations on pair of sites $AA$, $BB$, and $AB$. The results of MC simulations and extended pair approximation have indicated clearly that the concentration of cooperators on the sites of type $B$ is smaller than those we found on sites of type $A$ (shortly $\rho_B < \rho_A$) for the coexistence of $C$ and $D$ strategies. This latter success of pair approximation has inspired us to investigate the changes in the probabilities of one- and two-site configurations as well as in the transition rates between these configurations. Unfortunately, we were not able to reveal a simple and concise explanation on the mechanism supporting cooperation.

Besides the above discussed inhomogeneous dynamics there exists many other ways how inhomogeneous (and asymmetric) activity can be taken into consideration in the strategy adoption process. For example, we have studied what happens if the learning activity is reduced for the players of type $B$. In our notation it means that the strategy adoption probability (\ref{eq:Wtotal}) is modified by substituting $n_x$ for $n_y$ in the expression (\ref{eq:wobt}). Such a model can be used to study the effect of quenched spatial disorder on cooperation if the system consist of quick- ($A$) slow-witted ($B$) players. Surprisingly, MC simulations demonstrates that cooperation is not modified relevantly by the application of inhomogeneous learning activities on both lattice structures for $w=0.1$ and $\nu = 0.5$. The weak effect of this type of inhomogeneous learning activity is justified by the analytical techniques too. Namely, the extended pair approximation on square lattice (three-site approximation on the Kagome lattice) reproduces the result of homogeneous system. 

We have also studied a third type of models for which the quenched values of $w_{xy}=w_{yx}$ are chosen randomly to be $w$ or 1 with probabilities $\nu$ and $(1 - \nu)$. In this case the master-follower role is symmetric between the connected players and the dynamical inhomogeneity is represented by the randomly distributed reduced links. Similarly to the second case the simulations indicate no relevant changes in the strategy concentrations.

To have some further information about the role of the spatial inhomogeneities and the asymmetry in the imitation process we have studied a fourth case where players $A$ and $B$ are located periodically in checkerboard-like manner. In this case the role of reduced teaching from the sites of type $B$ is equivalent to the reduced learning activity at the sites of type $A$. Furthermore, the interactions are constrained to the $A-B$ pairs. The effect of this type of dynamical inhomogeneity on the strategy concentrations depends on the parameter values. An increase occurs in $\rho$ for low values of $b$. For sufficiently large $b$ values the concentration of cooperators (and also $b_{c2}$) decreases with $w$.

\begin{figure}[ht]
\centerline{\epsfig{file=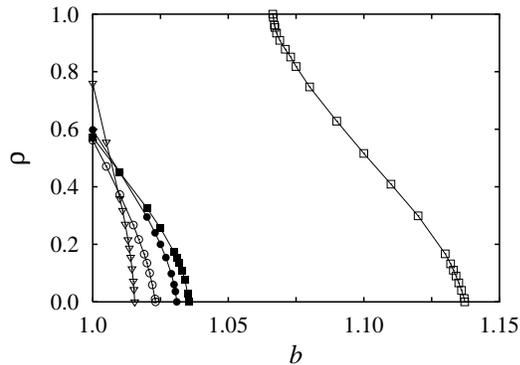,width=7cm}}
\caption{\label{fig:comp}Concentration of cooperators {\it vs.} $b$ on the square lattice for different systems: 
homogeneous (closed squares); reduced teaching (open squares) and learning (open circles) activity at players $B$ 
distributed randomly, checkerboard-like distribution of players $A$ and $B$ (triangles); and randomly distributed 
$w_{xy}=w_{yx}=0,w$ (closed circles) if $\nu=0.5$, $w=0.1$, and $K=1$.}
\end{figure}

In Figure \ref{fig:comp} we illustrate the results of $\rho$ deduced from MC data on the square lattice for the mentioned distributions of $w_{xy}$. The variations can be measured from the results obtained on the homogeneous system. It is clearly shown that the most relevant enhancement of cooperation is found when the teaching activity is reduced (or enhanced) from half of the players distributed randomly. According to preliminary results this is the only model among the investigated systems that is capable to maintain cooperation if $b > 1$ for the high noise limit.

In summary, we have studied two-strategy evolutionary PD games on two-dimensional lattices in which the master-follower asymmetry between two neighboring players is taken explicitly into account during the strategy adoption mechanism. Our analysis is focused on several cases of the quenched inhomogeneous dynamics. The most important increase of cooperativity is found for those systems where the enhanced teaching activity is concentrated on a portion of local players (masters or leaders) distributed randomly in the whole system. It is conjectured that the spontaneous appearance of local leaders can be recognized for those human and animal societies where mimics (learning) plays a crucial role in the maintenance of cooperation.

\begin{acknowledgments}
This work was supported by the Hungarian National Research Fund under 
Grant No. T-47003.
\end{acknowledgments}


\begin{thebibliography}{35}
\expandafter\ifx\csname natexlab\endcsname\relax\def\natexlab#1{#1}\fi
\expandafter\ifx\csname bibnamefont\endcsname\relax
  \def\bibnamefont#1{#1}\fi
\expandafter\ifx\csname bibfnamefont\endcsname\relax
  \def\bibfnamefont#1{#1}\fi
\expandafter\ifx\csname citenamefont\endcsname\relax
  \def\citenamefont#1{#1}\fi
\expandafter\ifx\csname url\endcsname\relax
  \def\url#1{\texttt{#1}}\fi
\expandafter\ifx\csname urlprefix\endcsname\relax\def\urlprefix{URL }\fi
\providecommand{\bibinfo}[2]{#2}
\providecommand{\eprint}[2][]{\url{#2}}

\bibitem[{\citenamefont{Weibull}(1995)}]{weibull_95}
\bibinfo{author}{\bibfnamefont{J.~W.} \bibnamefont{Weibull}},
  \emph{\bibinfo{title}{Evolutionary Game Theory}} (\bibinfo{publisher}{MIT
  Press}, \bibinfo{address}{Cambridge, MA}, \bibinfo{year}{1995}).

\bibitem[{\citenamefont{Gintis}(2000)}]{gintis_00}
\bibinfo{author}{\bibfnamefont{H.}~\bibnamefont{Gintis}},
  \emph{\bibinfo{title}{Game Theory Evolving}} (\bibinfo{publisher}{Princeton
  University Press}, \bibinfo{address}{Princeton}, \bibinfo{year}{2000}).

\bibitem[{\citenamefont{Cressman}(2003)}]{cressman_03}
\bibinfo{author}{\bibfnamefont{R.}~\bibnamefont{Cressman}},
  \emph{\bibinfo{title}{Evolutionary Dynamics and Extensive Form Games}}
  (\bibinfo{publisher}{MIT Press}, \bibinfo{address}{Cambridge, MA},
  \bibinfo{year}{2003}).

\bibitem[{\citenamefont{Hamilton}(1964)}]{hamilton_jtb64a}
\bibinfo{author}{\bibfnamefont{W.~D.} \bibnamefont{Hamilton}},
  \bibinfo{journal}{J. Theor. Biol.} \textbf{\bibinfo{volume}{7}},
  \bibinfo{pages}{1} (\bibinfo{year}{1964}).

\bibitem[{\citenamefont{Axelrod and Hamilton}(1981)}]{axelrod_s81}
\bibinfo{author}{\bibfnamefont{R.}~\bibnamefont{Axelrod}} \bibnamefont{and}
  \bibinfo{author}{\bibfnamefont{W.~D.} \bibnamefont{Hamilton}},
  \bibinfo{journal}{Science} \textbf{\bibinfo{volume}{211}},
  \bibinfo{pages}{1390} (\bibinfo{year}{1981}).

\bibitem[{\citenamefont{Fehr and G{\"a}chter}(2002)}]{fehr_n02}
\bibinfo{author}{\bibfnamefont{E.}~\bibnamefont{Fehr}} \bibnamefont{and}
  \bibinfo{author}{\bibfnamefont{S.}~\bibnamefont{G{\"a}chter}},
  \bibinfo{journal}{Nature} \textbf{\bibinfo{volume}{415}},
  \bibinfo{pages}{137} (\bibinfo{year}{2002}).

\bibitem[{\citenamefont{Hauert et~al.}(2002)\citenamefont{Hauert, De~Monte,
  Hofbauer, and Sigmund}}]{hauert_s02}
\bibinfo{author}{\bibfnamefont{C.}~\bibnamefont{Hauert}},
  \bibinfo{author}{\bibfnamefont{S.}~\bibnamefont{De~Monte}},
  \bibinfo{author}{\bibfnamefont{J.}~\bibnamefont{Hofbauer}}, \bibnamefont{and}
  \bibinfo{author}{\bibfnamefont{K.}~\bibnamefont{Sigmund}},
  \bibinfo{journal}{Science} \textbf{\bibinfo{volume}{296}},
  \bibinfo{pages}{1129} (\bibinfo{year}{2002}).

\bibitem[{\citenamefont{Nowak and May}(1993)}]{nowak_ijbc93}
\bibinfo{author}{\bibfnamefont{M.~A.} \bibnamefont{Nowak}} \bibnamefont{and}
  \bibinfo{author}{\bibfnamefont{R.~M.} \bibnamefont{May}},
  \bibinfo{journal}{Int. J. Bifurcat. Chaos} \textbf{\bibinfo{volume}{3}},
  \bibinfo{pages}{35} (\bibinfo{year}{1993}).

\bibitem[{\citenamefont{Nowak and Sigmund}(2004)}]{nowak_s04}
\bibinfo{author}{\bibfnamefont{M.~A.} \bibnamefont{Nowak}} \bibnamefont{and}
  \bibinfo{author}{\bibfnamefont{K.}~\bibnamefont{Sigmund}},
  \bibinfo{journal}{Science} \textbf{\bibinfo{volume}{303}},
  \bibinfo{pages}{793} (\bibinfo{year}{2004}).

\bibitem[{\citenamefont{Lindgren and Nordahl}(1994)}]{lindgren_pd94}
\bibinfo{author}{\bibfnamefont{K.}~\bibnamefont{Lindgren}} \bibnamefont{and}
  \bibinfo{author}{\bibfnamefont{M.~G.} \bibnamefont{Nordahl}},
  \bibinfo{journal}{Physica D} \textbf{\bibinfo{volume}{75}},
  \bibinfo{pages}{292} (\bibinfo{year}{1994}).

\bibitem[{\citenamefont{Nakamaru et~al.}(1997)\citenamefont{Nakamaru, Matsuda,
  and Iwasa}}]{nakamaru_jtb97}
\bibinfo{author}{\bibfnamefont{M.}~\bibnamefont{Nakamaru}},
  \bibinfo{author}{\bibfnamefont{H.}~\bibnamefont{Matsuda}}, \bibnamefont{and}
  \bibinfo{author}{\bibfnamefont{Y.}~\bibnamefont{Iwasa}}, \bibinfo{journal}{J.
  Theor. Biol.} \textbf{\bibinfo{volume}{184}}, \bibinfo{pages}{65}
  (\bibinfo{year}{1997}).

\bibitem[{\citenamefont{Szab{\'o} and F{\'a}th}(2006)}]{szabo_cm06}
\bibinfo{author}{\bibfnamefont{G.}~\bibnamefont{Szab{\'o}}} \bibnamefont{and}
  \bibinfo{author}{\bibfnamefont{G.}~\bibnamefont{F{\'a}th}}
  (\bibinfo{year}{2006}), \eprint{arXiv:cond-mat/0607344}.

\bibitem[{\citenamefont{Nowak et~al.}(1994)\citenamefont{Nowak, Bonhoeffer, and
  May}}]{nowak_ijbc94}
\bibinfo{author}{\bibfnamefont{M.~A.} \bibnamefont{Nowak}},
  \bibinfo{author}{\bibfnamefont{S.}~\bibnamefont{Bonhoeffer}},
  \bibnamefont{and} \bibinfo{author}{\bibfnamefont{R.~M.} \bibnamefont{May}},
  \bibinfo{journal}{Int. J. Bifurcat. Chaos} \textbf{\bibinfo{volume}{4}},
  \bibinfo{pages}{33} (\bibinfo{year}{1994}).

\bibitem[{\citenamefont{Vainstein and Arenzon}(2001)}]{vainstein_pre01}
\bibinfo{author}{\bibfnamefont{M.~H.} \bibnamefont{Vainstein}}
  \bibnamefont{and} \bibinfo{author}{\bibfnamefont{J.~J.}
  \bibnamefont{Arenzon}}, \bibinfo{journal}{Phys. Rev. E}
  \textbf{\bibinfo{volume}{64}}, \bibinfo{pages}{051905}
  (\bibinfo{year}{2001}).

\bibitem[{\citenamefont{Vukov and Szab{\'o}}(2005)}]{vukov_pre05}
\bibinfo{author}{\bibfnamefont{J.}~\bibnamefont{Vukov}} \bibnamefont{and}
  \bibinfo{author}{\bibfnamefont{G.}~\bibnamefont{Szab{\'o}}},
  \bibinfo{journal}{Phys. Rev. E} \textbf{\bibinfo{volume}{71}},
  \bibinfo{pages}{036133} (\bibinfo{year}{2005}).

\bibitem[{\citenamefont{Ebel and Bornholdt}(2002)}]{ebel_pre02}
\bibinfo{author}{\bibfnamefont{H.}~\bibnamefont{Ebel}} \bibnamefont{and}
  \bibinfo{author}{\bibfnamefont{S.}~\bibnamefont{Bornholdt}},
  \bibinfo{journal}{Phys. Rev. E} \textbf{\bibinfo{volume}{66}},
  \bibinfo{pages}{056118} (\bibinfo{year}{2002}).

\bibitem[{\citenamefont{Duran and Mulet}(2005)}]{duran_pd05}
\bibinfo{author}{\bibfnamefont{O.}~\bibnamefont{Duran}} \bibnamefont{and}
  \bibinfo{author}{\bibfnamefont{R.}~\bibnamefont{Mulet}},
  \bibinfo{journal}{Physica D} \textbf{\bibinfo{volume}{208}},
  \bibinfo{pages}{257} (\bibinfo{year}{2005}).

\bibitem[{\citenamefont{Abramson and Kuperman}(2001)}]{abramson_pre01}
\bibinfo{author}{\bibfnamefont{G.}~\bibnamefont{Abramson}} \bibnamefont{and}
  \bibinfo{author}{\bibfnamefont{M.}~\bibnamefont{Kuperman}},
  \bibinfo{journal}{Phys. Rev. E} \textbf{\bibinfo{volume}{63}},
  \bibinfo{pages}{030901(R)} (\bibinfo{year}{2001}).

\bibitem[{\citenamefont{Wu et~al.}(2006{\natexlab{a}})\citenamefont{Wu, Xu, and
  Wang}}]{wu_cpl06}
\bibinfo{author}{\bibfnamefont{Z.-X.} \bibnamefont{Wu}},
  \bibinfo{author}{\bibfnamefont{X.-J.} \bibnamefont{Xu}}, \bibnamefont{and}
  \bibinfo{author}{\bibfnamefont{Y.-H.} \bibnamefont{Wang}},
  \bibinfo{journal}{Chin. Phys. Lett.} \textbf{\bibinfo{volume}{23}},
  \bibinfo{pages}{531} (\bibinfo{year}{2006}{\natexlab{a}}).

\bibitem[{\citenamefont{Holme et~al.}(2003)\citenamefont{Holme, Trusina, Kim,
  and Minnhagen}}]{holme_pre03}
\bibinfo{author}{\bibfnamefont{P.}~\bibnamefont{Holme}},
  \bibinfo{author}{\bibfnamefont{A.}~\bibnamefont{Trusina}},
  \bibinfo{author}{\bibfnamefont{B.~J.} \bibnamefont{Kim}}, \bibnamefont{and}
  \bibinfo{author}{\bibfnamefont{P.}~\bibnamefont{Minnhagen}},
  \bibinfo{journal}{Phys. Rev. E} \textbf{\bibinfo{volume}{68}},
  \bibinfo{pages}{030901(R)} (\bibinfo{year}{2003}).

\bibitem[{\citenamefont{Vainstein et~al.}(2006)\citenamefont{Vainstein, Silva,
  and Arenzon}}]{vainstein_cm06}
\bibinfo{author}{\bibfnamefont{M.~H.} \bibnamefont{Vainstein}},
  \bibinfo{author}{\bibfnamefont{A.~T.~C.} \bibnamefont{Silva}},
  \bibnamefont{and} \bibinfo{author}{\bibfnamefont{J.~J.}
  \bibnamefont{Arenzon}} (\bibinfo{year}{2006}), \eprint{arXiv:Q-bio/0608009}.

\bibitem[{\citenamefont{Santos and Pacheco}(2005)}]{santos_prl05}
\bibinfo{author}{\bibfnamefont{F.~C.} \bibnamefont{Santos}} \bibnamefont{and}
  \bibinfo{author}{\bibfnamefont{J.~M.} \bibnamefont{Pacheco}},
  \bibinfo{journal}{Phys. Rev. Lett.} \textbf{\bibinfo{volume}{95}},
  \bibinfo{pages}{098104} (\bibinfo{year}{2005}).

\bibitem[{\citenamefont{Santos et~al.}(2006)\citenamefont{Santos, Rodrigues,
  and Pacheco}}]{santos_prslb06}
\bibinfo{author}{\bibfnamefont{F.~C.} \bibnamefont{Santos}},
  \bibinfo{author}{\bibfnamefont{J.~F.} \bibnamefont{Rodrigues}},
  \bibnamefont{and} \bibinfo{author}{\bibfnamefont{J.~M.}
  \bibnamefont{Pacheco}}, \bibinfo{journal}{Proc. Roy. Soc. Lond. B}
  \textbf{\bibinfo{volume}{273}}, \bibinfo{pages}{51} (\bibinfo{year}{2006}).

\bibitem[{\citenamefont{Kim et~al.}(2002)\citenamefont{Kim, Trusina, Holme,
  Minnhagen, Chung, and Choi}}]{kim_pre02}
\bibinfo{author}{\bibfnamefont{B.~J.} \bibnamefont{Kim}},
  \bibinfo{author}{\bibfnamefont{A.}~\bibnamefont{Trusina}},
  \bibinfo{author}{\bibfnamefont{P.}~\bibnamefont{Holme}},
  \bibinfo{author}{\bibfnamefont{P.}~\bibnamefont{Minnhagen}},
  \bibinfo{author}{\bibfnamefont{J.~S.} \bibnamefont{Chung}}, \bibnamefont{and}
  \bibinfo{author}{\bibfnamefont{M.~Y.} \bibnamefont{Choi}},
  \bibinfo{journal}{Phys. Rev. E} \textbf{\bibinfo{volume}{66}},
  \bibinfo{pages}{021907} (\bibinfo{year}{2002}).

\bibitem[{\citenamefont{Wu et~al.}(2006{\natexlab{b}})\citenamefont{Wu, Xu,
  Huang, Wang, and Wang}}]{wu_pre06}
\bibinfo{author}{\bibfnamefont{Z.-X.} \bibnamefont{Wu}},
  \bibinfo{author}{\bibfnamefont{X.-J.} \bibnamefont{Xu}},
  \bibinfo{author}{\bibfnamefont{Z.-G.} \bibnamefont{Huang}},
  \bibinfo{author}{\bibfnamefont{S.-J.} \bibnamefont{Wang}}, \bibnamefont{and}
  \bibinfo{author}{\bibfnamefont{Y.-H.} \bibnamefont{Wang}},
  \bibinfo{journal}{Phys. Rev. E} \textbf{\bibinfo{volume}{74}},
  \bibinfo{pages}{021107} (\bibinfo{year}{2006}{\natexlab{b}}).

\bibitem[{\citenamefont{Szab{\'o} et~al.}(2005)\citenamefont{Szab{\'o}, Vukov,
  and Szolnoki}}]{szabo_pre05}
\bibinfo{author}{\bibfnamefont{G.}~\bibnamefont{Szab{\'o}}},
  \bibinfo{author}{\bibfnamefont{J.}~\bibnamefont{Vukov}}, \bibnamefont{and}
  \bibinfo{author}{\bibfnamefont{A.}~\bibnamefont{Szolnoki}},
  \bibinfo{journal}{Phys. Rev. E} \textbf{\bibinfo{volume}{72}},
  \bibinfo{pages}{047107} (\bibinfo{year}{2005}).

\bibitem[{\citenamefont{Marro and Dickman}(1999)}]{marro_99}
\bibinfo{author}{\bibfnamefont{J.}~\bibnamefont{Marro}} \bibnamefont{and}
  \bibinfo{author}{\bibfnamefont{R.}~\bibnamefont{Dickman}},
  \emph{\bibinfo{title}{Nonequilibrium Phase Transitions in Lattice Models}}
  (\bibinfo{publisher}{Cambridge University Press},
  \bibinfo{address}{Cambridge}, \bibinfo{year}{1999}).

\bibitem[{\citenamefont{Dickman and Moreira}(1998)}]{dickman_pre98}
\bibinfo{author}{\bibfnamefont{R.}~\bibnamefont{Dickman}} \bibnamefont{and}
  \bibinfo{author}{\bibfnamefont{A.~G.} \bibnamefont{Moreira}},
  \bibinfo{journal}{Phys. Rev. E} \textbf{\bibinfo{volume}{57}},
  \bibinfo{pages}{1263} (\bibinfo{year}{1998}).

\bibitem[{\citenamefont{Hooyberghs et~al.}(2004)\citenamefont{Hooyberghs,
  Igl{\'o}i, and Vanderzande}}]{hooyberghs_pre04}
\bibinfo{author}{\bibfnamefont{J.}~\bibnamefont{Hooyberghs}},
  \bibinfo{author}{\bibfnamefont{F.}~\bibnamefont{Igl{\'o}i}},
  \bibnamefont{and}
  \bibinfo{author}{\bibfnamefont{C.}~\bibnamefont{Vanderzande}},
  \bibinfo{journal}{Phys. Rev. E} \textbf{\bibinfo{volume}{69}},
  \bibinfo{pages}{066140} (\bibinfo{year}{2004}).

\bibitem[{\citenamefont{Vukov et~al.}(2006)\citenamefont{Vukov, Szab{\'o}, and
  Szolnoki}}]{vukov_pre06}
\bibinfo{author}{\bibfnamefont{J.}~\bibnamefont{Vukov}},
  \bibinfo{author}{\bibfnamefont{G.}~\bibnamefont{Szab{\'o}}},
  \bibnamefont{and} \bibinfo{author}{\bibfnamefont{A.}~\bibnamefont{Szolnoki}},
  \bibinfo{journal}{Phys. Rev. E} \textbf{\bibinfo{volume}{73}},
  \bibinfo{pages}{067103} (\bibinfo{year}{2006}).

\bibitem[{\citenamefont{Perc}(2006)}]{perc_njp06}
\bibinfo{author}{\bibfnamefont{M.}~\bibnamefont{Perc}}, \bibinfo{journal}{New
  J. Phys.} \textbf{\bibinfo{volume}{8}}, \bibinfo{pages}{22}
  (\bibinfo{year}{2006}).

\bibitem[{\citenamefont{Traulsen et~al.}(2004)\citenamefont{Traulsen, R{\"o}hl,
  and Schuster}}]{traulsen_prl04}
\bibinfo{author}{\bibfnamefont{A.}~\bibnamefont{Traulsen}},
  \bibinfo{author}{\bibfnamefont{T.}~\bibnamefont{R{\"o}hl}}, \bibnamefont{and}
  \bibinfo{author}{\bibfnamefont{H.~G.} \bibnamefont{Schuster}},
  \bibinfo{journal}{Phys. Rev. Lett.} \textbf{\bibinfo{volume}{93}},
  \bibinfo{pages}{028701} (\bibinfo{year}{2004}).

\end{thebibliography}

\end{document}